\documentclass[12pt,preprint2]{emulateapj}

\usepackage{lineno}
\usepackage{natbib}
\usepackage{amsmath}
\usepackage{graphicx}
\usepackage{wasysym}
\usepackage{txfonts}
\usepackage{xspace}
\usepackage{url}
\usepackage{textcomp}
\usepackage{multirow}
\usepackage{lipsum}

\newcommand{\ks}{KS~1947+300\xspace}

\newcommand{\swift}{\textsl{Swift}\xspace}

\newcommand{\inte}{\textsl{INTEGRAL}\xspace}

\newcommand{\sax}{\textsl{BeppoSAX}\xspace}

\newcommand{\xte}{\textsl{RXTE}\xspace}
\newcommand{\nustar}{\textsl{NuSTAR}\xspace}

\newcommand{\snr}{S/N\xspace}

\newcommand{\redchi}{\ensuremath{\chi^{2}_\text{red}}\xspace}

\newcommand{\feka}{\ensuremath{\mathrm{Fe}~\mathrm{K}\alpha}\xspace}

\newcommand{\nh}{\ensuremath{{N}_\mathrm{H}}\xspace}

\shorttitle{\nustar discovery of a cyclotron line in \ks}
\shortauthors{F\"urst et al.}

\begin{document}

\title{\nustar discovery of a cyclotron line in \ks}

\author{Felix F\"urst\altaffilmark{1}}
\author{Katja Pottschmidt\altaffilmark{2,3}}
\author{J\"orn Wilms\altaffilmark{4}}
\author{Jamie Kennea\altaffilmark{5}}
\author{Matteo Bachetti\altaffilmark{6,7}}
\author{Eric Bellm\altaffilmark{1}}
\author{Steven E. Boggs\altaffilmark{8}}
\author{Deepto Chakrabarty\altaffilmark{9}}
\author{Finn E. Christensen\altaffilmark{10}}
\author{William W. Craig\altaffilmark{8}}
\author{Charles J. Hailey\altaffilmark{11}}
\author{Fiona Harrison\altaffilmark{1}}
\author{Daniel Stern\altaffilmark{12}}
\author{John A. Tomsick\altaffilmark{8}}
\author{Dominic J. Walton\altaffilmark{1}}
\author{William Zhang\altaffilmark{13}}

\altaffiltext{1}{Cahill Center for Astronomy and Astrophysics, California Institute of Technology, Pasadena, CA 91125, USA}
\altaffiltext{2}{Center for Space Science and Technology, University of Maryland
Baltimore County, Baltimore, MD 21250, USA}
\altaffiltext{3}{CRESST and NASA Goddard Space Flight Center, Astrophysics Science
Division, Code 661, Greenbelt, MD 20771, USA}
\altaffiltext{4}{Dr. Karl-Remeis-Sternwarte and ECAP, Sternwartstr. 7, 96049 Bamberg, Germany}
\altaffiltext{5}{Department of Astronomy \& Astrophysics, The Pennsylvania State University, University Park, PA 16802, USA}
\altaffiltext{6}{Universit\'e de Toulouse; UPS-OMP; IRAP; Toulouse, France}
\altaffiltext{7}{CNRS; Institut de Recherche en Astrophysique et Plan\'etologie, 31028 Toulouse cedex 4, France}
\altaffiltext{8}{Space Sciences Laboratory, University of California, Berkeley, CA 94720, USA}
\altaffiltext{9}{Kavli Institute for Astrophysics and Space Research, Massachusetts Institute of Technology, Cambridge, MA 02139, USA}
\altaffiltext{10}{DTU Space, National Space Institute, Technical University of Denmark, Elektrovej 327, 2800 Lyngby, Denmark}
\altaffiltext{11}{Columbia Astrophysics Laboratory, Columbia University, New York, NY 10027, USA}
\altaffiltext{12}{Jet Propulsion Laboratory, California Institute of Technology, Pasadena, CA 91109, USA}
\altaffiltext{13}{NASA Goddard Space Flight Center, Astrophysics Science
Division, Code 662, Greenbelt, MD 20771, USA}

\begin{abstract}
We present a spectral analysis of three simultaneous \nustar and \swift/XRT observations of the transient Be-neutron star binary \ks taken during its  outburst in 2013/2014. These broad-band observations were supported by \swift/XRT monitoring snap-shots every 3\,days, which we use to study the evolution of the spectrum over the outburst. We find strong changes of the power-law photon index, which shows a weak trend of softening with increasing X-ray flux.  The neutron star shows very strong pulsations with a period of $P\approx18.8$\,s.  The 0.8--79\,keV broad-band spectrum can be described by a power-law with an exponential cutoff and a black-body component at low energies. 
During the second observation we detect a cyclotron resonant scattering feature at 12.5\,keV, which is absent in the phase-averaged spectra of observations 1 and 3. Pulse phase-resolved spectroscopy reveals that the strength of the feature changes strongly with pulse phase and is most prominent during the broad minimum of the pulse profile. At the same phases the line  also becomes visible in the first and third observation at the same energy. This discovery implies that  \ks  has a  magnetic field strength of $B\approx1.1\times10^{12}(1+z)$\,G, which is at the lower end of known cyclotron line sources. 
\end{abstract}

\keywords{accretion, accretion disks --- radiation: dynamics --- stars: neutron --- X-rays: binaries --- X-rays: individual (KS 1947+300)}

\section{Introduction}
\ks was independently discovered with \textsl{Mir-Kvant}/TTM by \citet{borozdin90a} and with \textsl{CGRO}/BATSE by \citet{finger94a} and \citet{chakrabarty95a} during outbursts in 1989 and 1994, respectively. \citet{swank00a} used \xte data during an outburst in 2000 and the 18.7\,s pulse period to identify both detections as the same accreting neutron star. The optical companion was identified by \citet{negueruela03a} as a Be-type star at a distance of $\sim$10\,kpc, assuming a standard luminosity. \citet{galloway04a} determined the orbit and found an orbital period of $P_\text{orb}=41.5$\,d, with a very low eccentricity of  $e=0.034\pm0.007$. 

In 2000 \xte performed an extensive campaign to monitor a large outburst that reached a peak flux of 120\,mCrab in the 1.5-12\,keV band. \citet{galloway04a} found that the energy spectrum could be described with a simple Comptonization model \citep[\texttt{compTT},][]{titarchuk94a, hua95a}, a model often applied to highly magnetized neutron stars. They found no source-intrinsic absorption, but a broad excess around 10\,keV which they described with a hot black-body component with $kT_\text{bb}=3\text{--}4$\,keV. 

Using \sax data taken during the decay of the same major outburst, \citet{naik06a} found a similar spectral shape but a much cooler black-body component, $kT_\text{bb}\approx0.6$\,keV. They additionally found evidence for a \feka line at $\sim$6.6\,keV. 

The major outburst was followed by a series of weaker outbursts, the strongest of which occurred in 2004 April and reached  $\sim$45\,mCrab  in the  1.5--12\,keV energy band. This series of outbursts was serendipitously  monitored by \inte during its Galactic Plane scans. \citet{tsygankov05a} described the \inte/ISGRI and JEM-X spectra using a power-law with a high-energy cut-off and found indications for a spectral softening with increased flux. 

Accreting neutron-stars sometime show cyclotron resonant scattering features (CRSFs) in their hard X-ray spectra. These absorption-like lines are the only way to directly measure the magnetic field strength close to the neutron star surface. They are produced by photons that scatter off electrons quantized onto Landau-levels in the strong magnetic field ($B\approx10^{12}$\,G) of the neutron star. Their energy is directly related to the strength of the magnetic field in the line forming region via the ``12-B-12''-rule:
\begin{equation}
 E_\text{CRSF}=11.57\times{B}_\text{12}(1+z)\,\text{keV}
 \end{equation}
where $B_{12}$ is the magnetic field in $10^{12}$\,G and $z$ the gravitational redshift \citep[for a detailed discussion see, e.g.,][]{schoenherr07a}. 
Theoretically CRSF could also result in emission features \citep{schoenherr07a}, but there is only little observational evidence to date \citep[a possible detection was reported for 4U~1626$-$67, see][]{iwakiri12a}.
Despite coverage with \xte, \sax, and \inte, a CRSF was not detected in previous outbursts of \ks \citep{naik06a, galloway04a, tsygankov05a}. 

\begin{deluxetable}{lcccc}
\tablecolumns{5}
\tablewidth{0pc}
\tablecaption{Observation log for the three simultaneous observations.\label{tab:obslog}}
\tablehead{\colhead{Observatory/} &\colhead{ ObsID} &\colhead{start date} &\colhead{exposure} & \colhead{pulse period}  \\
\colhead {Instrument} & \colhead{} & \colhead{MJD (d)} & \colhead {(ks)} &  \colhead{(s)}}
\startdata
  \nustar &  80002015002 & 56586.79 & 18.4  & $18.80584(16)$ \\
 \swift/XRT & 00032990003 & 56587.25 & 0.37  & \\
  \nustar &  80002015004 & 56618.91 & 18.6  & $18.78399(7)$ \\
 \swift/XRT & 00032990013 & 56618.94 & 0.93  & \\
  \nustar &  80002015006 & 56635.75 &  25.4 & $18.77088(6)$ \\
 \swift/XRT & 00032990020 &  56635.67 &  0.91 &
\enddata
\end{deluxetable}

\begin{figure}
\centering
\includegraphics[width=0.99\columnwidth]{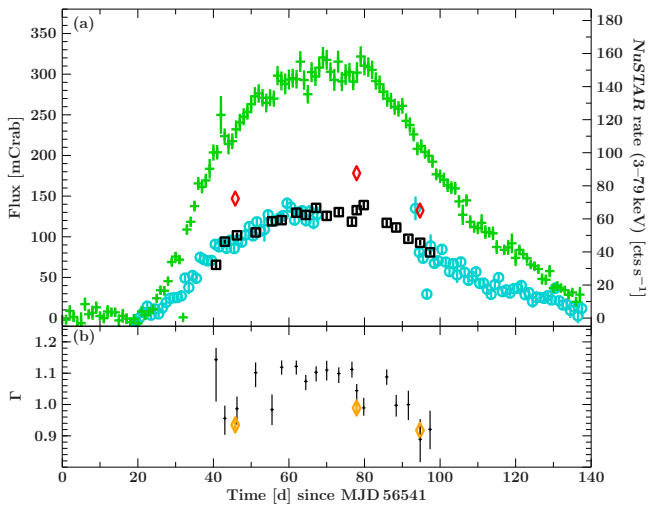}
\caption{\textit{(a)} Lightcurve of the 2013 outburst. Data from \swift/BAT \citep[15--50\,keV,][]{swiftbatref} are shown as green crosses, MAXI (2--20\,keV) as blue circles, \swift/XRT (0.5--8\,keV) as black squares and \nustar/FPMA (3--79\,keV) as red diamonds. All data are scaled to the Crab count-rate in the respective energy band. The \nustar count-rate for the \nustar data is shown on the right-hand $y$-axis. \textit{(b)} Best-fit power-law index $\Gamma$ of the \swift/XRT spectra, for details of the model see text. The orange diamonds  show combined XRT and \nustar results.
}
\label{fig:batlc}
\end{figure}

\ks has been in quiescence from 2004--2013. In 2013 October MAXI \citep{maxiref} detected increased flux levels \citep{Atel5438}. The beginning of an outburst was immediately confirmed by \swift/XRT \citep{Atel5441} and monitored by \swift/BAT. We triggered  \swift/XRT $\sim$1\,ks snap-shot observations every 3\,days to monitor the outburst in soft X-rays (Figure~\ref{fig:batlc}). It reached a peak flux of $\sim$130\,mCrab in the 3--10\,keV energy band, very comparable to the maximum of the bright 2000 outburst \citep{naik06a}. Additionally, we triggered three observations with the \textsl{Nuclear Spectroscopy Telescope Array} \citep[\nustar;][]{harrison13a}. 
An overview of the observations and their exposure times can be found in Table~\ref{tab:obslog}.

\section{Observations \& data reduction}
\label{sec:data}
\subsection{\nustar}
\label{susec:nustar}
\nustar consists of two independent grazing incidence telescopes focusing X-rays between 3--79\,keV  onto two Focal Plane Modules, FPMA and FPMB. We used the standard NUSTARDAS software v1.2.0 as distributed with HEASOFT~6.14 to extract spectra and lightcurves. \mbox{\nustar} spectra were used between 3--60\,keV. Above 60\,keV the calibration at the time of writing shows  increased systematic uncertainties, and we therefore do not use those data. A detailed analysis of the high energy calibration will be presented in a forthcoming publication. The source data were extracted from a $130''$ radius circular region centered at $\alpha_\text{J2000}=19^\mathrm{h}49^\mathrm{m}36^\mathrm{s}$ and $\delta_\text{J2000}=+30^\circ12'22''$. 
Background spectra were extracted from a circular region with $105''$ radius as far away from the source as possible. This formally introduces systematic uncertainties in the background estimation,
 but since \ks is at least a factor of 40 brighter than the background at all energies, the influence on the source flux is negligible.  Light-curves were extracted with a resolution of 1\,s, the resolution corresponding to dead-time measurements in the standard operating mode.

\subsection{\swift/XRT}
\label{susec:xrt}
Data from the \swift/XRT \citep{swiftxrtref} were extracted following the standard guidelines\footnote{\url{http://www.swift.ac.uk/analysis/xrt/}}, using XSELECT to extract spectra and lightcurves and \texttt{xrtmkarf} to create the response files. All data were obtained in window timing mode. The source data were extracted from a circular region with a radius of 20 sky pixels ($\approx47''$). Background spectra were extracted from the wings of the PSF using an annular region between 90  and 110 pixels radius ($212''$ and $259''$, respectively). In XRT \ks is a factor of 50 brighter than the background at all energies, rendering small uncertainties in the background negligible. We used the XRT spectra in the energy range between 0.8--10\,keV. At lower energies the windowed timing mode shows larger calibration uncertainties and we therefore decided not to use those energies.\footnote{see \url{http://www.swift.ac.uk/analysis/xrt/digest\_cal.php\#abs}}

\subsection{Reduction methods}
All timing information for both satellites was transferred to the solar barycenter, using the FTOOL \texttt{barycorr} and the DE-200 solar system ephemeris \citep{solarDE200}, and corrected for the binary motion using the ephemeris by \citet{galloway04a}. Timing and spectral analysis was performed using the Interactive Spectral Interpretation System \citep[ISIS v1.6.2, ][]{houck00a}. All uncertainties are given at the 90\% level ($\Delta\chi^2=2.7$ for one parameter  of interest), unless otherwise noted.

\section{Phase-averaged spectroscopy}
\label{sec:phasavg}

For spectral modeling, we use FPMA and FPMB spectra as well as the corresponding XRT data for each epoch, as detailed in Table~\ref{tab:obslog}. The  X-ray continuum is very well described with a simple power-law with an exponential cutoff (model \texttt{cutoffpl} in XSPEC) plus a black-body. The black-body is responsible for about 50\%  of the flux at 2\,keV and follows the overall flux evolution of the data. It likely originates  from the hot-spot of the neutron star surface.
The \texttt{compTT} model used by \citet{naik06a} and \citet{galloway04a} results in a clearly worse fit. 

\citet{naik06a} and \citet{galloway04a} measured an absorption column towards the source which was lower than the maximal Galactic value along that line of sight \citep[$\sim9\times10^{21}$\,cm$^{-2}$, ][]{kalberla05a}. We therefore allow the absorption to vary in our model, but require it to be the same in all three observations. We describe it using an updated version of the \texttt{tbabs} \citep{wilms00a} model\footnote{\url{http://pulsar.sternwarte.uni-erlangen.de/wilms/research/tbabs/}}, with the corresponding abundances and cross-sections by \citet{verner96a}. 
Our best fit value of $8.45\pm0.20\times10^{21}$\,cm$^{-2}$ is consistent with the 21\,cm value along the line of sight and also with the \nh obtained from the reddening of the source \citep[$A_V=3.38$;][]{negueruela03a} when using the calibration of \citet{predehl95a} as updated by \citet{nowak12a}.

\begin{figure}
\centering
\includegraphics[width=0.99\columnwidth]{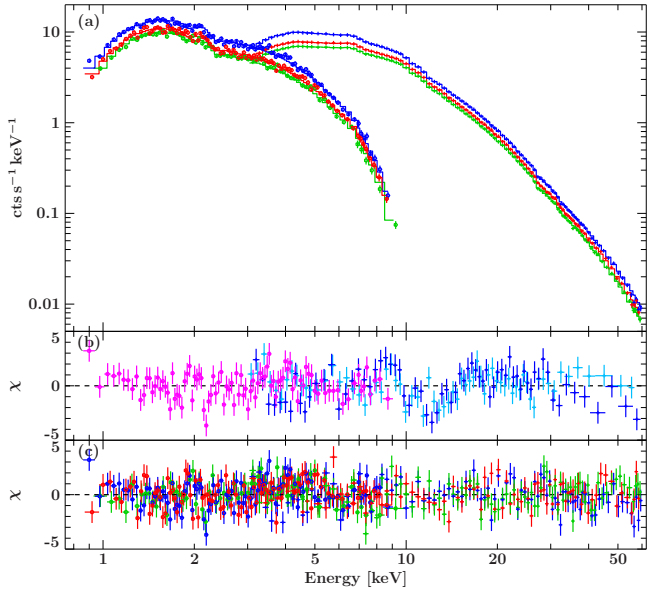}
\caption{\textit{(a)} Data and best-fit model of \swift/XRT (circles) and \nustar/FPMA (crosses) for all three epochs. Observation 1 is shown in red, observation 2 in blue, and observation 3 in green. \textit{(b)} Residuals of observation 2 to the best-fit model without a CRSF, rebinned for plotting purposes; \swift/XRT is shown in magenta, \nustar/FPMA in blue and FPMB in light blue. \textit{(c)} Residuals to the best-fit model including the CRSF of all three observations in terms of $\chi$.
}
\label{fig:3spec}
\end{figure}


Adding a Gaussian \feka line at around 6.5\,keV to the continuum model we obtain a good description of the spectra, with a $\chi^2=3369.9$ for 3068 d.o.f. ($\redchi=1.10$). The data are shown in Figure~\ref{fig:3spec}\textit{(a)}. However, close inspection of the residuals of the second observation reveals significant residuals around 13\,keV (see Figure~\ref{fig:3spec}\textit{(b)}). We therefore add a multiplicative absorption line with a Gaussian optical depth profile (model \texttt{gabs} in XSPEC) to the model for the second observation. This additional component improves the fit significantly to $\chi^2=3284.6$ for 3063 d.o.f. \citep[$\redchi=1.07$, $F$-test false positive probability $1.5\times10^{-15}$, after][]{datarederroana}. This model is shown in Figure~\ref{fig:3spec}(a), the best-fit residuals in Figure~\ref{fig:3spec}(c), and its parameters are given in Table~\ref{tab:bestfit_all}. The fluxes are given in \nustar/FPMA normalization and we allow for small cross-calibration differences to \swift/XRT and FPMB using the multiplicative factors $\text{C}_\text{XRT}$ and $\text{C}_\text{FPMB}$, respectively.

\begin{deluxetable}{rlll}
\tablewidth{0pc}
\tablecaption{Fit Parameters for the Best-fit Phase-averaged Model.\label{tab:bestfit_all}}
\tablehead{\colhead{Parameter} & \colhead{Obs. I} & \colhead{Obs. II} & \colhead{Obs. III}}
\startdata
 $ N_\text{H}\tablenotemark{a}\tablenotemark{f} $ & $0.848^{+0.024}_{-0.020}$ & $0.848^{+0.024}_{-0.020}$ & $0.848^{+0.024}_{-0.020}$ \\
 $ \mathcal{F}_\text{1--60\,keV}\tablenotemark{b}$ &  $7.202\pm0.027$ & $9.08\pm0.04$ & $6.491\pm0.021$ \\
 $ {L}_\text{1--60\,keV}\tablenotemark{c}$ & $8.62 \pm 0.03$ &  $10.87 \pm 0.05$ &  $7.768 \pm 0.024$ \\
 $ \Gamma$ & $0.928\pm0.014$ & $0.982^{+0.022}_{-0.015}$ & $0.927^{+0.014}_{-0.012}$ \\
 $ E_\text{cut} (\text{keV})$ & $22.5^{+0.5}_{-0.4}$ & $24.2^{+0.6}_{-0.5}$ & $22.13^{+0.45}_{-0.24}$ \\
 $ E_\text{CRSF} (\text{keV})\tablenotemark{f}$ & $12.2^{+0.5}_{-0.7}$ & $12.2^{+0.5}_{-0.7}$ & $12.2^{+0.5}_{-0.7}$ \\
 $ \sigma_\text{CRSF} (\text{keV})\tablenotemark{f}$ & $2.5^{+1.3}_{-0.6}$ & $2.5^{+1.3}_{-0.6}$ & $2.5^{+1.3}_{-0.6}$ \\
 $ d_\text{CSRF} (\text{keV})$ & $0.00^{+0.04}_{-0.05}$ & $0.16^{+0.15}_{-0.05}$ & $-0.05^{+0.04}_{-0.10}$ \\
 $ A_\text{BB}\tablenotemark{c}$ & $4.12^{+0.21}_{-0.20}$ & $5.73^{+0.28}_{-0.52}$ & $3.45^{+0.19}_{-0.18}$ \\
 $ kT (\text{keV})$ & $0.663^{+0.017}_{-0.018}$ & $0.745^{+0.017}_{-0.026}$ & $0.636^{+0.023}_{-0.016}$ \\
 $ A(\text{Fe\,K}\alpha)\tablenotemark{d}$ & $1.78^{+0.20}_{-0.19}$ & $2.28^{+0.22}_{-0.25}$ & $1.20^{+0.16}_{-0.12}$ \\
 $ E(\text{Fe\,K}\alpha) (\text{keV})\tablenotemark{e}$ & $6.575^{+0.030}_{-0.032}$ & $6.563^{+0.031}_{-0.028}$ & $6.539^{+0.029}_{-0.031}$ \\
 $ \sigma(\text{Fe\,K}\alpha)(\text{keV})\tablenotemark{e}$ & $0.29\pm0.05$ & $0.31\pm0.04$ & $0.25\pm0.04$ \\
 $ \text{C}_\text{XRT}$ & $0.967\pm0.014$ & $0.966^{+0.012}_{-0.015}$ & $0.982^{+0.017}_{-0.015}$ \\
 $ \text{C}_\text{FPMB}$ & $1.0207\pm0.0023$ & $1.0308\pm0.0021$ & $1.0232\pm0.0021$
\enddata
\tablenotetext{a}{ in $10^{22}$\,cm$^{-2}$}
\tablenotetext{b}{ unabsorbed flux in $10^{-9}$\,erg\,s$^{-1}$\,cm$^{-2}$}
\tablenotetext{c}{ luminosity for a distance of 10\,kpc in $10^{37}$\,erg\,s$^{-1}$}
\tablenotetext{c}{ in $10^{36}$\,erg\,s$^{-1}$ for a distance of 10\,kpc}
\tablenotetext{d}{ in $10^{-3}$\,ph\,s$^{-1}$\,cm$^{-2}$}
\tablenotetext{e}{ in keV}
\tablenotetext{f}{parameter tied across observations}

\end{deluxetable}

We search for similar absorption lines in the spectra of the other two observations. For that we require the energy and width of the \texttt{gabs} component to be the same in all observations and allow only the depth to vary in a simultaneous fit to all three data-sets. In both other data-sets the line is not significantly detected (Table~\ref{tab:bestfit_all}). The 90\% uncertainties are clearly below the depth of the line during  observation 2, indicating a physical change in the source spectrum over the outburst.

\subsection{Time-resolved spectral analysis}
\label{susec:tresspec}
To study the evolution of the spectrum over the outburst we use all available \swift/XRT data between MJD\,56581--56639, and describe them with the same \texttt{cutoffpl} plus \texttt{bbody} model as the time-averaged spectra. 
We fix the cutoff-energy at 23.2\,keV, the average value in the \nustar data and the absorption column to $8.45\times10^{21}$\,cm$^{-2}$, since the Galactic absorption column should not vary.

We find a strong degeneracy between the power-law slope and the black-body temperature due to the limited energy range covered by \swift/XRT. From the simultaneous \nustar and \swift/XRT spectra it becomes clear, however, that an almost linear correlation between the black-body temperature and the unabsorbed 3--10\,keV flux is present, 
\begin{equation}
kT_\text{bbody}=m\times\mathcal{F}_\text{3--10\,keV}\text{,}
\end{equation}
where $kT_\text{bbody}$ is measured in keV and $\mathcal{F}_\text{3--10\,keV}$ in keV\,s$^{-1}$\,cm$^{-2}$.
In the simultaneous fits we find $m= 0.576\pm0.010$\,s\,cm$^{2}$.
 We use this correlation to tie the black-body temperature to the X-ray flux in the time-resolved XRT spectra and consequently replace the degeneracy with an empirically motivated correlation.
 
The remaining free parameters in the model are the photon-index of the power-law, the 3--10\,keV flux and the relative normalization of the black-body component. The latter does not show significant changes with time. The model describes all 19 XRT spectra relatively well, with an average \redchi=1.05 for 429 d.o.f. 

As shown in Figure~\ref{fig:batlc}(b), the photon-index is highly variable and seems to soften with increased X-ray flux. This correlation is marginally significant at a bit above the 1$\sigma$ level, with Spearman's rank correlation coefficient $\rho=0.31$. It also becomes clear that all three \nustar observations were performed during phases with relatively hard spectra. \citet{tsygankov05a} found a similar correlation between then photon index and the X-ray luminosity in the harder \xte energy band (3--100\,keV).

\section{Phase-resolved spectroscopy}
\label{sec:phasres}
\begin{figure*}
\centering
\includegraphics[width=0.99\textwidth]{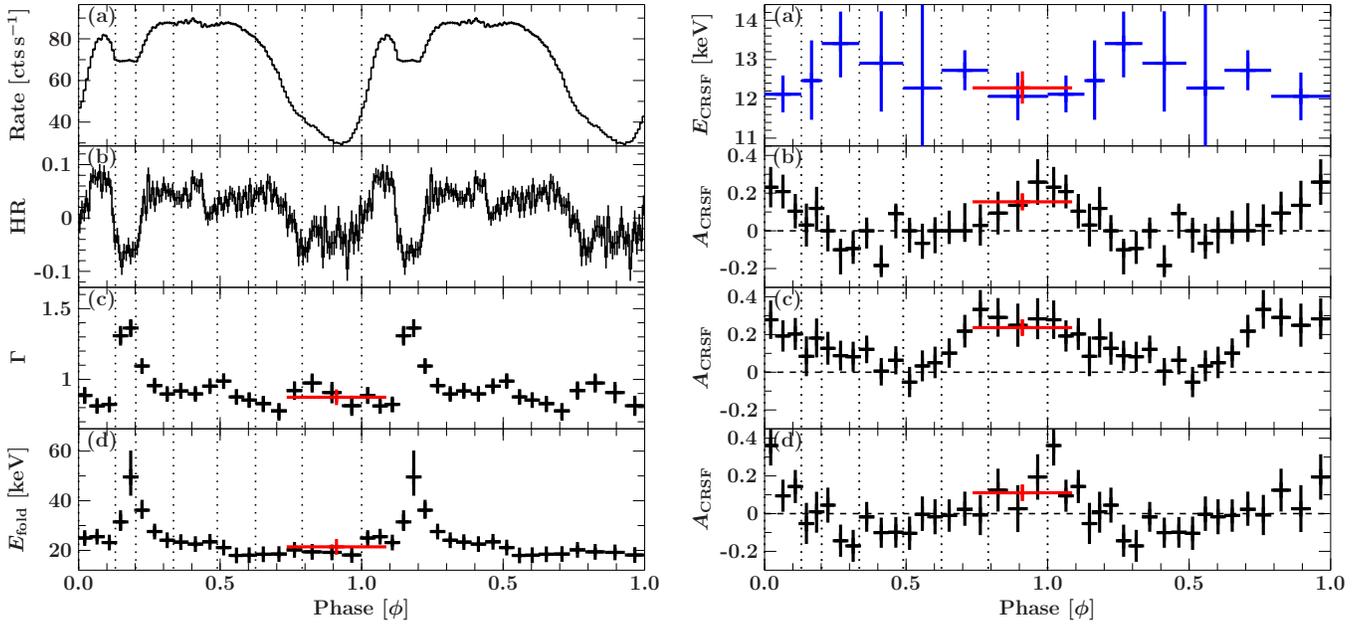}
\caption{Results of the phase resolved spectroscopy. The left panel shows for observation 2 the \textit{(a)} pulse-profile between 3--79\,keV, \textit{(b)} hardness ratio $(H-S)/(H+S)$ between the energy bands $S$=4.5--6\,keV and $H$=10--15\,keV. The dotted lines indicate the seven, wider phase bins used to measure the CRSF energy. The photon-index $\Gamma$ and the folding energy are shown in panels \textit{(c)} and \textit{(d)}, respectively. The right panel shows the parameters of the CRSF: \textit{(a)} line energy for the second observation, \textit{(b), (c)}, and \textit{(d)} line strength for the first, second, and third observation, respectively. The red data-points show the results of the phase bin covering the phases with the most significant CRSF in observation 2; see text for details. The pulse is repeated once for clarity. 
}
\label{fig:phaseres}
\end{figure*}

To investigate changes with viewing angle  onto the neutron star we split each \nustar data set into 20 phase bins. For this analysis we did not use the \swift/XRT data, as their short exposure does not allow us to split them up further. We define the phase bins to stretch over intervals of similar flux and hardness ratio, see Figure~\ref{fig:phaseres}. The pulse profile changes drastically with energy, developing from one broad pulse to a double-peaked profile, with a narrow primary  and broader secondary peak above $\sim$25\,keV \citep[see also][]{naik06a}.

To define the phase bins individually for each observation, we first measure the local pulse period by folding the cleaned \nustar event list on trial periods around the expected period of 18.8\,s, following the description given by \citet{leahy98a}. The uncertainties are estimated by phase-connecting pulse-profiles from the beginning and end of each observation. We do not allow for a change in the pulse period during one observation, but the error introduced is below the precision needed for the analysis presented here. 
The measured periods 
show a continuous spin-up over the duration of the outburst (see Table~\ref{tab:obslog}), in agreement with the \swift/XRT snap-shots and the \textsl{Fermi}/GBM pulsar monitoring\footnote{\url{http://gammaray.nsstc.nasa.gov/gbm/science/pulsars/lightcurves/ks1947.html}} \citep{finger09a}. 

To describe the phase-resolved spectra we use the same  model as for the phase-averaged spectrum, but fix the line energies and widths of the CRSF and the \feka line as well as the temperature of the black-body due to the reduced statistical quality of the spectra. 
This model results in a very good description of the data in all phase bins, with an average $\chi^2_\text{red}=1.02$ for 345 d.o.f.

 Figure~\ref{fig:phaseres} shows 
 the continuum parameters only for the second observation, since it provides the best statistics, but the other two observations show very similar behavior.  Both the photon-index $\Gamma$ and the folding energy $E_\text{fold}$ show a very strong dependence on phase, confirming the results by \citet{naik06a} at a much higher resolution in phase space (Figure \ref{fig:phaseres} \textit{left, (c)} and \textit{(d)}). Between phase 0.1--0.2 we observe a strong increase in $\Gamma$ and $E_\text{fold}$, coincident with the small dip between the narrow first peak and the broad main peak. 

The strength of the CRSF  shows a very interesting behavior with pulse phase, as shown for all observations in Figure~\ref{fig:phaseres}, right panel. As expected from the phase-averaged spectra, the line is clearly strongest in observation 2, being detectable over a wide phase-range between phases 0.6--1.3. During the main peak of the pulse-profile the line strength drops to 0. In observation 1 the line is also significantly detected in absorption between phases 0.9--1.1. Around phase 0.3 there are low significance indications that the line is instead visible in emission.
In observation 3 the line strength is scattering around 0, only one phase bin around phase 1.0 shows an absorption line clearly above the 95\% limit.
%

In the phase-resolved spectra we allow the iron line normalization and the black-body normalization to vary to obtain a statistically acceptable fit (not shown in Figure~\ref{fig:phaseres}). We carefully checked that any variation in these parameters does not influence the strength of the CRSF. While the iron line shows the largest equivalent width during the minimum of the pulse profile, i.e., at the same phases where the  CRSF is most prominent, it does not influence the spectral shape at the CRSF energy. The black-body did not vary significantly over the pulse phase. 

To investigate the energy dependence of the CRSF with pulse-phase, we extracted spectra from observation 2 using seven, wider phase bins to increase the \snr.   For these spectra we also allow the iron line energy and width, as well as the blackbody temperature to vary. We keep the width of the CRSF fixed to the best phase-averaged value, as it otherwise became unconstrained  during the fits. As can be seen in Figure~\ref{fig:phaseres}, \textit{right (a)} we do not detect a significant variation of the line energy with pulse phase. Between phases 0.4--0.6 we again detected no significant line, resulting in an unconstrained energy. 

Besides changes with pulse-phase, changes of the line energy with luminosity are quite common  \citep[see, e.g.][among others]{caballero07a, tsygankov10a, fuerst14a}. To search for such a luminosity dependence between observations, we extract spectra for each observation of those phases, in which the line was significantly detected  in observation 2.
This approach allows to obtain the most significant line and therefore most precise energy measurement, as indicated by the blue data points in Figure~\ref{fig:phaseres}.
We describe the spectra with the same model as for the seven wide phase bins described above. We do not detect a significant change of the line energy, with the measured values being $12.5\pm0.7$\,keV, $12.3\pm0.5$\,keV, and  $12.2\pm0.9$\,keV for observation 1, 2, and 3 respectively. 

\section{Summary \& Discussion}
\label{sec:outlook}
We have presented a spectral analysis of three \nustar observations of the Be-X-ray binary \ks with simultaneous \swift/XRT data, taken during its large 2013/2014  outburst. The broad spectral coverage provided by the combination of these two instruments 
allowed us to discover a CRSF absorption feature around 12.5\,keV. The feature was significantly detected  in the phase-averaged spectrum of the brightest observation, and during the broad pulse minimum in phase-resolved spectroscopy in all observations. During the pulse maximum the feature is not seen significantly, either in absorption or emission.

The line energy and width  is similar to the lines detected in 4U\,0115+63 and  Swift~J1626.6$-$5156 \citep{white83a, decesar13a}. 
We deduce a surface magnetic field of $\sim1.1\times10^{12}(1+z)$\,G, assuming that the line is the fundamental line. Here $z$ is the gravitational redshift, defined by
\begin{equation}
(1+z)^{-1}=\sqrt{1-\frac{2GM}{Rc^2}}.
\end{equation}
For typical neutron star parameters, $z\approx0.3$  if the line-forming region is close to the surface. 
This magnetic field strength  puts \ks at the lower end of known cyclotron lines sources \citep[cf.][]{caballero07a}.


During the broad minimum phase of the pulse profile, we detect the CRSF in all three observations.  
The luminosity near $10^{38}$\,erg\,s$^{-1}$ puts \ks clearly in the super-critical accretion regime, where the radiation pressure is strong enough to decelerate the in-falling matter before the neutron star surface via a radiation-dominated shock \citep{becker12a}. In this regime, a negative correlation between the CRSF energy and luminosity is expected \citep{becker12a}, as observed, for example, in V\,0332+53 \citep{tsygankov10a}.  If the correlation were of a similar strength as observed in V\,0332+53 we would not have detected it due to the very small range of luminosities sampled.

The time-resolved \swift/XRT spectra show a strongly variable photon-index $\Gamma$ over the outburst, with changes of 10\% or more within 3\,days and softening with increasing X-ray flux. This softening agrees with the expected behavior in the supercritical accretion regime, as shown by \citet{klochkov11a} for various other sources. However, because we restricted the model to describe basically all changes in spectral hardness in the photon-index, it is probable that the true physical changes are more complex than a variable photon-index, e.g., the black-body temperature might vary independently of the X-ray flux.  Nonetheless, intrinsic source variability must be present. 

We clearly detect a \feka line in all data sets, with an energy significantly above the line energy for neutral iron (see Table~\ref{tab:bestfit_all}) and broadened in excess of the energy resolution of \nustar. While Doppler-broadening could be responsible for part of the observed width, the increased energy indicates that the fluorescence region is slightly ionized and the observed broadening originates from a blend of \feka at different low ionization states. The data do not allow us to disentangle different lines from one single broad line.


\acknowledgments
We would like to thank Matthias K\"uhnel, Ralf Ballhausen,  Fritz Schwarm, and Peter Kretschmar for useful discussions.
This work was supported under NASA Contract No.~NNG08FD60C, and
made use of data from the \nustar mission, a project led by
the California Institute of Technology, managed by the Jet Propulsion
Laboratory, and funded by the National Aeronautics and Space
Administration. We thank the \mbox{\nustar} Operations, Software and
Calibration teams for support with the execution and analysis of
these observations. This research has made use of the \nustar
Data Analysis Software (NuSTARDAS) jointly developed by the ASI
Science Data Center (ASDC, Italy) and the California Institute of
Technology (USA). 
This research has made use of ISIS functions provided by ECAP/Remeis observatory
and MIT (http://www.sternwarte.uni-erlangen.de/isis/).
We thank the Deutsches Zentrum f\"ur Luft- und Raumfahrt for partial support under DLR grant 50 OR 1113.
We would like to thank the anonymous referee for the useful comments.

{\it Facilities:} \facility{NuSTAR}, \facility{Swift}

%



\end{document}